\documentclass[twocolumn]{autart}

\usepackage{graphics}
\usepackage{stfloats}
\usepackage{epsfig}
\usepackage{amssymb}
\usepackage{amsmath}
\usepackage{epstopdf}
\usepackage{color}
\usepackage{enumerate}
\usepackage{subfigure}
\usepackage{cite}
\usepackage{algorithm}
\usepackage{algpseudocode}
\renewcommand{\algorithmicrequire}{\textbf{Input:}}  
\usepackage{flushend}
\DeclareMathOperator*{\minimize}{minimize}

\newcommand{\addRed}[1]{\textcolor{red}{#1}}

\newtheorem{mythm}{Theorem}

\newtheorem{remark}{Remark}
\newtheorem{myassum}{Assumption}

\begin{document}

\begin{frontmatter}

\title{Data-Driven Min-Max MPC for Linear Systems:\\ Robustness and Adaptation\thanksref{footnoteinfo}}
\thanks[footnoteinfo]{F. Allg{o}wer is thankful that his work was funded by Deutsche Forschungsgemeinschaft (DFG, German Research
Foundation) under Germany’s Excellence Strategy - EXC 2075 - 390740016 and under grant 468094890.
F. Allg\"{o}wer acknowledges the support by the Stuttgart
Center for Simulation Science (SimTech).
The authors thank the International Max Planck Research School
for Intelligent Systems (IMPRS-IS) for supporting Yifan Xie.}

\author[Stuttgart]{Yifan Xie},
\author[Stuttgart]{Julian Berberich},
\author[Stuttgart]{Frank Allg{\"o}wer},

\address[Stuttgart]{University of Stuttgart, Institute for Systems Theory and Automatic Control, 70550 Stuttgart, Germany, \\
Email: \{yifan.xie, julian.berberich, frank.allgower@ist.uni-stuttgart.de\}}


\begin{abstract}
Data-driven controllers design is an important research problem, in particular when data is corrupted by the noise.
In this paper, we propose a data-driven min-max model predictive control (MPC) scheme using noisy input-state data for unknown linear time-invariant (LTI) system.
The unknown system matrices are characterized by a set-membership representation using the noisy input-state data.
Leveraging this representation, we derive an upper bound on the worst-case cost and determine the corresponding optimal state-feedback control law through a semidefinite program (SDP).
We prove that the resulting closed-loop system is robustly stabilized and satisfies the input and state constraints.
Further, we propose an adaptive data-driven min-max MPC scheme which exploits additional online input-state data to improve closed-loop performance.
Numerical examples show the effectiveness of the proposed methods.
\end{abstract}

\end{frontmatter}

\section{Introduction}\label{sec:1}
Data-driven system analysis and control have received increasing interest in the recent years.
Unlike the model-based control methods, which depend on prior knowledge of system models identified from measured data using system identification methods \cite{ljung1999system} or derived from first principles, data-driven control approaches design controllers directly from the available data.
Several works have focused on designing controllers for unknown linear time-invariant (LTI) systems directly from noisy data \cite{van2020noisy,van2020data, berberich2023combining, persis2020formulas, berberich2020robust, bisoffi2021trade, van2023quadratic}, assuming assumptions on the noise such as energy bounds or instantaneous bounds.
A matrix inequality can effectively characterize a set of LTI systems that explain the measured data, forming the basis of the data informativity framework \cite{van2020data,van2023informativity}.
Within this framework, data-driven controller designs aim to find a controller that stabilizes all systems that are consistent with the data.
Various controller design methods have been proposed, including $H_2$ and $H_\infty$ control \cite{van2020noisy, van2023quadratic, berberich2023combining}, linear quadratic regulator approaches \cite{persis2020formulas}, and stabilization \cite{berberich2020robust, persis2020formulas, van2020noisy}.
Nevertheless, \addRed{the controller design with input and state constraints using noisy data remains largely unexplored in this framework.}
In this paper, we propose data-driven min-max model predictive control (MPC) schemes in this framework to design a controller that robustly stabilizes the system and handles ellipsoidal input and state constraints.

MPC is widely used due to its ability to handle constraints and consider performance criteria \cite{rawlings2017model}.
The fundamental concept of MPC is to solve an open-loop optimal control problem at each sampling time, which uses the system dynamics to predict future open-loop trajectories.
Recently, data-driven MPC approaches have been studied, which directly use the measured input-output data to predict the future outputs \cite{yang2015data, coulson2019data,berberich2021guarantees, bongard2022robust, markovsky2021review, faulwasser2023behavioral,markovsky2023datapower,VERHEIJEN2023handbook}.
This data-driven MPC framework is based on the Fundamental Lemma \cite{willems2005note, markovsky2008data},
which states that for a controllable LTI system, all possible system trajectories can be parameterized in terms of linear combinations of time-shifts of one persistently exciting trajectory.
This framework requires the availability of persistently exciting data, enabling the unique representation of the system from the data in the noise-free scenarios.
In case of bounded output measurement noise, robust data-driven MPC schemes have been developed, which guarantee practical stability for the closed-loop system \cite{coulson2019data,berberich2021guarantees}.
These MPC schemes can be expanded via suitable constraint tightening to ensure robust state or output constraint satisfaction in the presence of bounded process or measurement noise \cite{kloppelt2022novel,berberich2020robustcon}.
While these MPC schemes provide strong theoretical guarantees for guaranteeing constraint satisfaction of unknown systems based only on measured data, the employed constraint tightenings suffer from possibly large conservatism.

Min-max MPC can effectively address scenarios involving parametric uncertainty on the system dynamics and disturbance, as discussed in \cite{kothare1996robust,bemporad2003minmax, diehl2007formulation, lu2000quasi, limon2006input, lazar2008input, lazar2009further}.
The basic idea of min-max MPC is to design control inputs that minimize the worst-case cost w.r.t. disturbances and/or parametric uncertainty in order to robustly stabilize the system.
An especially popular approach is to employ linear matrix inequalities (LMIs) \cite{scherer2000linear, boyd1994linear} in the min-max MPC framework \cite{kothare1996robust} to obtain a tractable state-feedback control law.
This approach involves solving an LMI-based optimization problem at each time step that incorporates constraints and a description of the parametric uncertainty, thereby guaranteeing robust stability.
The min-max MPC schemes typically require prior knowledge of the parametric uncertainty set, i.e., a known polytopic set \cite{kothare1996robust, lu2000quasi}.
However, addressing min-max MPC schemes without prior knowledge of the parametric uncertainty set, relying solely on available data, remains an open challenge.

In this work, we propose a data-driven min-max MPC framework to control LTI systems with unknown system matrices and additive process noise using noisy input-state data.
Our approach relies on a representation of the system matrices consistent with a sequence of noisy input-state data by using a quadratic matrix inequality \cite{berberich2023combining, bisoffi2021trade}.
The scheme involves an infinite-horizon cost as well as ellipsoidal input and state constraints.
It can be interpreted as a time-varying $H_2$ state-feedback controller design, analogous to the model-based min-max MPC scheme in \cite{kothare1996robust}.
We show that the proposed data-driven min-max MPC guarantees closed-loop recursive feasibility, constraint satisfaction and robust stability.
Further, we propose an adaptive data-driven min-max MPC scheme that integrates online collected input-state data.
Utilizing these online data reduces the parametric uncertainty on the system dynamics, thus the closed-loop performance resulting from the adaptive data-driven min-max MPC scheme improves.
Numerical examples shows that
the proposed scheme ensures robust stability and constraint satisfaction in a less conservative fashion than the data-driven MPC based on the Fundamental Lemma \cite{berberich2020robustcon}.

We note that the recent works \cite{hu2024robust, nguyen2023lmirobust} also propose data-driven MPC schemes for linear systems using ideas from the data informativity framework.
However, in these papers, the data are assumed to be noise-free, contrary to our framework which allows for process noise in the data.
In \cite{li2024data}, a data-driven MPC scheme employing noisy data is proposed, focusing on the $H_\infty$ control objective.
However, they assume an energy bound on the online noise, implying that the noise convergences to zero for time approach infinity, and establish closed-loop stability accordingly.
In contrast, we address a more practical scenario with instantaneous noise bound and establish robust stability for the closed-loop system.
In the recent work \cite{rotulo2022online}, an online data-driven approach is proposed for iteratively learning controllers for systems with dynamic changes over time.
The data considered is noise-free and constraints are not taken into account.
Further, the literature contains various approaches on model-based adaptive MPC schemes \cite{lorenzen2019update,lorenzen2017adaptive,kohler2019,kohler2021robust,sasfi2023robust,zhang2020adaptive,tanaskovic2013adaptive, heirung2017dual,zhu2019constrained,zhang2019adaptive}.
Specifically, adaptive tube-based MPC schemes aim to construct prediction tubes for robust constraint satisfaction and incorporate model adaptation using set-membership estimation \cite{lorenzen2019update,kohler2019,kohler2021robust,zhang2020adaptive,sasfi2023robust,lorenzen2017adaptive}.
Model-based adaptive min-max MPC schemes update the parametric uncertainty set at each time step and design a MPC controller to robustly stabilize the uncertain systems \cite{zhang2019adaptive,zhu2019constrained}, which assumes prior knowledge of the parametric uncertainty set.
In contrast, the proposed approach relies on an ellipsoidal uncertainty characterization based on the recent data-driven control literature \cite{berberich2023combining, bisoffi2021trade}.
This allows to use LMI methods for the design of a state-feedback-based MPC scheme with robust stability guarantees.

The remainder of this paper is organized as follows.
Section~\ref{sec:2} introduces necessary preliminaries about the data-driven parameterization and the problem setup.
In Section~\ref{sec:3}, we propose a data-driven min-max MPC problem with input and state constraints.
We prove recursive feasibility, constraint satisfaction and robust stability for the closed-loop system.
In Section~\ref{sec:4}, we consider an adaptive data-driven min-max MPC scheme which uses online data to reduce uncertainty and improve performance.
We illustrate the advantage of the proposed schemes with numerical examples in Section~\ref{sec:5}.
Finally, we conclude the paper in Section~\ref{sec:6}.
Preliminary results on data-driven min-max MPC were presented in the conference paper \cite{xie2024minmax}.
The present paper extends \cite{xie2024minmax} in multiple directions.
First, \cite{xie2024minmax} assumes that the online measurements used for feedback are noise-free, whereas we consider noise in both offline and online data.
As a result, the theoretical analysis in the present paper is more involved and provides robust stability guarantees of a robust positive invariant (RPI) set around the origin.
On the contrary, the noise-free setup in \cite{xie2024minmax} allowed to prove exponential stability.
Further, the present paper proposes an adaptive data-driven min-max MPC scheme which, as shown with a numerical example, can substantially reduce conservatism using online data.

\emph{Notation:} Let $\mathbb{I}_{[a, b]}$ denote the set of integers in the interval $[a, b]$,  $\mathbb{I}_{\geq 0}$ denote the set of nonnegative integers, and $\mathbb{I}_{[a, +\infty)}$ denote the set of integers larger than or equal to $a$.
For a matrix $P$, we write $P\succ 0$ if $P$ is positive definite and $P\succeq 0$ if $P$ is positive semi-definite.
For a vector $x$ and a matrix $P\succ 0$, we write $\|x\|_P=\sqrt{x^\top Px}$.
\addRed{
For a matrix $P=P^\top$, we denote by $\lambda_{\min}(P)$ the minimal eigenvalue of the matrix $P$.}

\section{Preliminaries}\label{sec:2}
In Section \ref{sec:2.1}, the considered problem setup is introduced.
In Section \ref{sec:2.2}, we present the data-driven system parameterization used for the proposed min-max MPC approach.

\subsection{Problem Setup}\label{sec:2.1}
In this paper, we consider an unknown discrete-time LTI system
\begin{equation}\label{system}
x_{t+1}=A_sx_t+B_su_t+\omega_t,
\end{equation}
where $x_t\in\mathbb{R}^n$ denotes the state, $u_t\in\mathbb{R}^m$ denotes the input, and $\omega_t\in\mathbb{R}^n$ denotes the unknown noise for $t\in\mathbb{I}_{\geq 0}$.
The matrices $A_s\in\mathbb{R}^{n\times n}$ and $B_s\in\mathbb{R}^{n\times m}$ are assumed to be unknown.
The noise $\omega_t$ is assumed to satisfy the following assumption.

\begin{myassum}\label{assumption1}\upshape
For all $t\in \mathbb{I}_{\geq 0}$, the noise $\omega_t\in\mathbb{R}^n$ satisfies \addRed{$\|\omega_t\|_G\leq 1$
for a known matrix $G\succ 0$}.
\end{myassum}

\addRed{When $G=\epsilon^{-2}I$, Assumption~\ref{assumption1} implies $\omega_t^\top\omega_t\leq \epsilon^2$. Similar noise bounds are common in set-membership identification  \cite{hjalmarsson1993discussion} and data-driven controller design from noisy data \cite{bisoffi2021trade,berberich2023combining}.}

We define a sequence of offline input, noise and corresponding state of length $T_f$ from the system \eqref{system}, which are denoted in the matrices
\begin{equation}\nonumber
\begin{aligned}
U_f&:=\begin{bmatrix}u_0^f & u_1^f &\ldots &u_{T_f-1}^f\end{bmatrix},\\
W_f&:=\begin{bmatrix}\omega_0^f &\omega_1^f &\ldots &\omega_{T_f-1}^f\end{bmatrix},\\
X_f&:=\begin{bmatrix}x_0^f & x_1^f &\ldots &x_{T_f}^f\end{bmatrix}.
\end{aligned}
\end{equation}
Throughout this paper, we assume that the offline input-state measurements $U_f$ and $X_f$ are available. The noise \addRed{sequence} $W_f$ is unknown, but every element in $W_f$ satisfies the bound in Assumption \ref{assumption1}.
In Section \ref{sec:3}, the data-driven min-max MPC algorithm only based on the knowledge of offline input-state measurements $(U_f, X_f)$.
Moreover, in Section \ref{sec:4}, we employ online input-state measurements in addition to $(U_f, X_f)$ to design an adaptive min-max MPC algorithm.

Our objective is to stabilize the origin for the unknown LTI system \eqref{system}, while the closed-loop input and state satisfy given constraints.
We consider the origin for simplicity but note that the results in this paper can be adapted for non-zero equilibria.
In order to stabilize the origin, we define the following quadratic stage cost function
\begin{equation}\nonumber
l(u, x)=\|u\|_R^2+\|x\|_Q^2,
\end{equation}
where $R, Q\succ 0$.
We consider ellipsoidal constraints on the input and the state, i.e.,
\begin{subequations}\label{constraints}
\begin{align}
&\|u_t\|_{S_u}\leq 1, \forall t\in \mathbb{I}_{\geq 0},\label{constraints:input}\\
&\|x_t\|_{S_x}\leq 1, \forall t\in \mathbb{I}_{\geq 0},\label{constraints:state}
\end{align}
\end{subequations}
where $S_u\succ 0$ and  $S_x\succeq 0$.
\addRed{This generalizes the ellipsoidal constraints with $S_u=S_x=I$ commonly used in LMI-based min-max MPC \cite{kothare1996robust}.
Polytopic constraints can be addressed by deriving an inner ellipsoidal constraint set from the given polytopic constraint set and then apply the proposed method to ensure constraint satisfaction.}

\subsection{Data-driven Parameterization}\label{sec:2.2}
Given that the matrices $A_s$ and  $B_s$ are unknown, the knowledge about the system relies on inferring information from input-state measurements.
In this section, we introduce the employed data-driven parameterization method using offline input and state measurements.

First, we define the set of system matrices $(A, B)$ consistent with the offline data $x_i^f, u_i^f, x_{i+1}^f, i\in\mathbb{I}_{[0,T_f-1]}$ by
\begin{equation}\nonumber
\Sigma_i^f\!\!:=\!\!\left\{\!(A, B)\!:\!
\begin{gathered}
    \eqref{system} \text{ holds for some }\omega_i^f\\
\text{ satisfying Assumption~\ref{assumption1}}
\end{gathered}\right\}.
\end{equation}
This set includes all system matrices for which there exists a noise realization satisfying Assumption \ref{assumption1} and the system dynamics \eqref{system}.
We proceed analogous to  \cite{van2020noisy, bisoffi2021trade, berberich2023combining} to derive a data-driven parametrization of the system matrices.
Using the system dynamics \eqref{system}, the state $x_i^f, x_{i+1}^f$ and input $u_i^f$ satisfy the following equation
\[\omega_i^f=x_{i+1}^f-A_sx_i^f-B_su_i^f.\]
Thus, the set $\Sigma_i^f$ can be equivalently characterized by the following quadratic matrix inequality
\begin{equation}\label{sigma_f}
\Sigma_i^f\!\!=\!\!\left\{\!\!\!\!\begin{gathered}
(A, B):\\
    \begin{bmatrix}
    I\\ A^\top \\B^\top
\end{bmatrix}^\top\!\!\!
\begin{bmatrix}
I &x_{i+1}^f\\
0 &-x_i^f\\
0 &-u_i^f
\end{bmatrix}\!\!\!
\begin{bmatrix}
\addRed{G^{-1}} &0\\
0 &-I
\end{bmatrix}\!\!\!
\begin{bmatrix}
I &x_{i+1}^f\\
0 &-x_i^f\\
0 &-u_i^f
\end{bmatrix}^\top\!\!\!
\begin{bmatrix}
    I\\ A^\top \\B^\top
\end{bmatrix}\!\!\succeq\!\! 0
\end{gathered}\!\right\}.
\end{equation}

Furthermore, the set of system matrices consistent with the sequence of offline input-state measurements $( U_f,X_f)$ is defined by\begin{equation}\nonumber
\mathcal{C}_f:=\bigcap_{i=0}^{T_f-1}\Sigma_i^f.
\end{equation}
We can characterize $\mathcal{C}_f$ by the following quadratic matrix inequality \cite{berberich2023combining, bisoffi2021trade}
\begin{equation}\label{C}
\mathcal{C}_f\!\!=\!\!\!\left\{\!\!(A, B):\!\!\!\!\!\!\!
\begin{gathered}
\begin{bmatrix}
I &A &B
\end{bmatrix}
\Pi_f(\tau)
\begin{bmatrix}
I &A &B
\end{bmatrix}^\top\!\succeq\! 0,  \\
\forall \tau\!\!=\!\!(\tau_0, \ldots, \tau_{T_f-1}), \tau_i\!\geq\! 0, i\!\in\!\mathbb{I}_{[0, T_f-1]}\!
\end{gathered}\!
\right\},
\end{equation}
where
\begin{equation}\label{Pi}
\Pi_f(\tau)=\sum_{i=0}^{T_f-1}\tau_i
\begin{bmatrix}
I &x_{i+1}^f\\
0 &-x_i^f\\
0 &-u_i^f
\end{bmatrix}
\!\!
\begin{bmatrix}
\addRed{G^{-1}} &0\\
0 &-I
\end{bmatrix}
\begin{bmatrix}
I &x_{i+1}^f\\
0 &-x_i^f\\
0 &-u_i^f
\end{bmatrix}^\top.
\end{equation}

We will later use the data-driven parameterization of $\mathcal{C}_f$ in equation \eqref{C} to formulate the data-driven min-max MPC problem, thus designing a controller that robustly stabilizes all system with matrices in $\mathcal{C}_f$.

\section{Data-driven Min-Max MPC}\label{sec:3}
In this section, we present a data-driven min-max MPC problem with input and state constraints using offline input-state measurements.
We restrict the optimization to state-feedback control laws, which allows to reformulate the data-driven min-max MPC problem as an SDP.
We establish that the resulting MPC approach is recursively feasible and the closed-loop system is robustly stabilized and satisfies the input and state constraints.

\subsection{Data-driven Min-Max MPC Problem}\label{sec:3.1}

At time $t$, given offline input-state measurements $(U_f, X_f)$ and an initial state $x_t$, the data-driven min-max MPC optimization problem is formulated as follows:
\begin{subequations}\label{mpc:robust}
\begin{align}
J_\infty^*(x_t):=&\min_{\bar{u}(t)}\max_{(A, B)\in\mathcal{C}_f}\sum_{k=0}^{\infty}l(\bar{u}_k(t), \bar{x}_k(t))\label{mpc:robust_obj}\\
\text{s.t.}\quad &\bar{x}_{k+1}(t)=A\bar{x}_k(t)+B\bar{u}_k(t),\label{mpc:robust_con1}\\
&\bar{x}_0(t)=x_t.\label{mpc:robust_con2}\\
&\addRed{\|\bar{u}_k(t)\|_{S_u}\leq 1, \forall k\in\mathbb{I}_{\geq 0},}\label{mpc:robust_con3}\\
&\addRed{\|\bar{x}_k(t)\|_{S_x}\leq 1, \forall (A, B)\in\mathcal{C}_f, k\in\mathbb{I}_{\geq 0},}\label{mpc:robust_con4}
\end{align}
\end{subequations}
The objective function aims to minimize the worst-case value of the sum of infinite stage cost
among all consistent system matrices in $\mathcal{C}_f$ by adapting the control input $\bar{u}_k(t), \forall k\in\mathbb{I}_{\geq 0}$.
In the optimization problem, $\bar{x}_{k}(t)$ and $\bar{u}_k(t)$ are the predicted state and control input at time $t+k$ based on the measurement at time $t$.
\addRed{We use the nominal system dynamics \eqref{mpc:robust_con1} with $(A, B)\in\mathcal{C}_f$ for future state prediction in constraint \eqref{mpc:robust_con1}.}
In constraint \eqref{mpc:robust_con2}, we initialize $\bar{x}_0(t)$ as the state measurement at time $t$.
In constraints \eqref{mpc:robust_con3} and \eqref{mpc:robust_con4}, \addRed{the predicted input and state satisfy the ellipsoidal constraints in \eqref{constraints} for any nominal system dynamics \eqref{mpc:robust_con1} with $(A, B)\in\mathcal{C}_f$.}

\vspace{-8pt}
\begin{remark}\upshape
In the data-driven min-max MPC problem \eqref{mpc:robust}, we use the nominal system dynamics without the noise $\omega_t$ to predict the future state.
This approach avoids an additional maximization with regard to the noise in the min-max MPC problem.
\addRed{The constraints are imposed on the predicted input and state.}
Even though the influence of the noise is not considered in the min-max problem, the proposed method guarantees robust stability \addRed{and constraint satisfaction} for the closed-loop system in the presence of noise (cf. Section \ref{sec:3.3}).
\end{remark}
\vspace{-8pt}
\begin{remark}\upshape
Similar to the existing LMI-based min-max MPC scheme in \cite{kothare1996robust}, we consider to minimize the worst-case value of infinite-horizon cost.
The infinite-horizon cost allows us to  reformulate the data-driven min-max MPC problem as an SDP.
This reformulation fits well to the data-driven system parametrization described by \eqref{C}, as detailed in Section \ref{sec:3.2}.
\end{remark}

\subsection{Reformulation based on LMIs}\label{sec:3.2}
The data-driven min-max MPC problem \eqref{mpc:robust} is intractable because of the min-max formulation and the constraints for all possible $(A,B)$ within a set characterized by data.
To effectively address problem \eqref{mpc:robust} and derive a tractable solution, we limit our focus to find a \addRed{time-varying} state-feedback control law of the form $\bar{u}_k(t)=\addRed{F_t}\bar{x}_k(t)$, where $\addRed{F_t}\in\mathbb{R}^{m\times n}$.
In the following, we formulate an SDP to derive an upper bound on the optimal cost over the set $\mathcal{C}_f$ and to determine a state-feedback gain that minimizes this upper bound.

At time $t$, given offline input-state measurements $(U_f, X_f)$, an initial state $x_t\in\mathbb{R}^n$ and a constant $c>\lambda_{\min}(Q)$, the SDP is formulated as follows:
\begin{subequations}\label{sdp:robust}
\begin{align}
    &\minimize\limits_{\gamma>0, H\in\mathbb{R}^{n\times n}, L\in\mathbb{R}^{m\times n}, \tau\in\mathbb{R}^{T_f}}\gamma\label{sdp:robust_obj}\\
    \text{s.t. }
    &\begin{bmatrix}1 &x_t^\top\\
x_t &H\end{bmatrix}\succeq 0, \label{sdp:robust_con1}\\
    &\begin{bmatrix}
        \begin{bmatrix}
            -H+\frac{\gamma}{c} I &0\\
            0 &0
        \end{bmatrix}+\Pi_f(\tau) &
        \begin{bmatrix}
            0\\
            H\\
            L
        \end{bmatrix}
        & 0\\
        \begin{bmatrix}
            0 &H &L^\top
        \end{bmatrix} &-H &\Phi^\top\\
        0 &\Phi & -\gamma I
    \end{bmatrix}\prec 0, \label{sdp:robust_con2}\\
    &\tau=(\tau_0, \ldots, \tau_{T_f-1}), \tau_i\geq 0, \forall i\in\mathbb{I}_{[0, T_f-1]},\label{sdp:robust_con3}\\
    &\begin{bmatrix}
        H &L^\top\\
        L &S_u^{-1}
    \end{bmatrix}\succeq 0,\label{sdp:robust_con_input}\\
    &\begin{bmatrix}
        H &H\\
        H &S_x^{-1}
    \end{bmatrix}\succeq 0.\label{sdp:robust_con_state}
\end{align}
\end{subequations}

where $\Phi=\begin{bmatrix}M_R L\\ M_Q H\end{bmatrix}$, $M_R^\top M_R=R$ and $M_Q^\top M_Q=Q$.
\addRed{The optimal solution of problem \eqref{sdp:robust} depends on the current measured state $x_t$.}
\addRed{For a given state $x_t$}, the optimal solution of problem \eqref{sdp:robust}  is denoted by \addRed{$\gamma_{x_t}^\star, H_{x_t}^\star, L_{x_t}^\star, \tau_{x_t}^\star$}.
The corresponding optimal state-feedback gain is given by \addRed{$F_{x_t}^\star=L_{x_t}^\star (H_{x_t}^\star)^{-1}$} \addRed{and we define $P_{x_t}^\star=\gamma_{x_t}^\star (H_{x_t}^\star)^{-1}$.}

\begin{remark}\upshape
The constant $c$ is required to prove robust stability for the resulting closed-loop system, as detailed in Section \ref{sec:3.3}.
At the initial time $t=0$, we select $c$ such that the optimization problem \eqref{sdp:robust} is feasible for the initial state $x_0$.
This approach ensures convexity and recursive feasibility of the SDP problem \eqref{sdp:robust}, thereby simplifying computational burden.
The precise choice of $c$ influences the performance.
As will be proved later in Theorem \ref{theorem2}, $\|x_t\|_P^2$ is lower bounded by $\|x_t\|_Q^2$ and upper bounded by $c\|x_t\|^2$.
Therefore, this requires $c\geq \lambda_{\min}(Q)$.
A smaller value of $c$ gives a smaller robust positive invariant (RPI) set to which the closed loop converges, as shown later in Theorem \ref{theorem2}, but also a smaller feasible region and possibly larger optimal cost \addRed{$\gamma^\star_{x_t}$}.
\end{remark}

\addRed{
\begin{remark}\upshape
Problem \eqref{sdp:robust} provides an upper bound on the optimal cost of the problem \eqref{mpc:robust} and a corresponding state-feedback control gain.
This will be shown in Theorem~1 below.
Additionally, as will be established in Theorem \ref{theorem2}, the constraints \eqref{sdp:robust_con_input} and \eqref{sdp:robust_con_state} ensure that the input and state constraints in \eqref{constraints} are satisfied for the closed-loop system and the derived state-feedback gain.
\end{remark}
}

We solve the SDP problem \eqref{sdp:robust} in a receding-horizon manner to repeatedly find an optimal state-feedback gain, see Algorithm~1.
In particular, at time $t=0$, we solve the optimization problem \eqref{sdp:robust} and \addRed{implement only the first computed input $u_t=F_{x_t}^\star x_t$.
We then update $t=t+1$, measure the state at this next time step, and solve the optimization problem \eqref{sdp:robust} again.
If the optimal solution $\gamma_{x_t}^\star>\frac{c^2}{\lambda_{\min}(Q)\lambda_{\min}(G)}$, we implement the first computed input $u_t=F_{x_t}^\star x_t$ and repeat this procedure.
If the optimal solution $\gamma_{x_t}^\star\leq\frac{c^2}{\lambda_{\min}(Q)\lambda_{\min}(G)}$, then we define  $\tilde{F}=F_{x_{t-1}}^\star$ and $\tilde{P} = P_{x_{t-1}}^\star$.
From this time onward, we stop solving the problem \eqref{sdp:robust} and directly apply $\tilde{F}$ to the system.
}
\addRed{Algorithm~1 returns a dual-mode controller. In a neighborhood of the origin, a static state-feedback control law is implemented.
Outside this neighborhood, receding horizon control is employed, similar to the robust dual-mode MPC controller proposed in \cite{michalska1993robust}.}


\begin{algorithm}[htb]
\begin{algorithmic}[1]
\caption{\!Data-driven min-max MPC scheme.\!\!\!}
    \State \algorithmicrequire{ $U_f, X_f$, $Q, R, S_x, S_u$, $c$, \addRed{$G$}}\;
    \State At time $t=0$, measure state $x_0$\;
    \State Solve the problem \eqref{sdp:robust}\;
    \State Apply the input $u_t=\addRed{F_{x_t}^\star} x_t$\;
    \State Set $t=t+1$,  measure state $x_t$\;
    \State Solve the problem \eqref{sdp:robust}\;
    \addRed{
    \If{$\gamma_{x_t}^\star>\frac{c^2}{\lambda_{\min}(Q)\lambda_{\min}(G)}$}\;
    \State Apply the input $u_t=F_{x_t}^\star x_t$\;
    \State Set $t=t+1$,  measure state $x_t$, go back to 6\;
    \ElsIf{$\gamma_{x_t}^\star\leq\frac{c^2}{\lambda_{\min}(Q)\lambda_{\min}(G)}$}
    \State Set $\tilde{F}=F_{x_{t-1}}^\star, \tilde{P} = P_{x_{t-1}}^\star$\;
    \State Apply the input $u_t=\tilde{F} x_t$\;
    \State Set $t=t+1$,  measure state $x_t$, go back to 12\;
    \EndIf}
    \label{algorithm:robust}
\end{algorithmic}
\end{algorithm}

In the following theorem, we first neglect the input and state constraints \eqref{mpc:robust_con3}-\eqref{mpc:robust_con4} in the data-driven min-max MPC problem \eqref{mpc:robust}.
We show that the optimal cost of problem \eqref{sdp:robust} is an upper bound on the optimal cost of \eqref{mpc:robust_obj}-\eqref{mpc:robust_con2} using \eqref{sdp:robust_con1}-\eqref{sdp:robust_con3}.
Later in Theorem \ref{theorem2}, we will show that Algorithm~1 ensures constraint satisfaction in closed loop.

\begin{mythm}
\label{theorem1}\upshape
Given a state $x_t\in\mathbb{R}^n$ at time $t$ and a constant $c> \lambda_{\min}(Q)$, suppose there exist $\gamma$, $H$, $ L$, $\tau$ such that the LMIs \eqref{sdp:robust_con1}-\eqref{sdp:robust_con3} hold.
Let $P=\gamma H^{-1}$.
Then, the optimal cost of \eqref{mpc:robust_obj}-\eqref{mpc:robust_con2} is guaranteed to be at most $\|x_t\|_{P}^2$ and $\|x_t\|_{P}^2$ is upper bounded by $\gamma$, i.e.,
\[J^\star_\infty (x_t)\leq \|x_t\|_{P}^2\leq\gamma.\]
\end{mythm}
\begin{pf}
Applying the Schur complement to the constraint \eqref{sdp:robust_con2} twice yields the equivalent inequalities
\begin{subequations}\label{M1}
\begin{align}
    &\begin{bmatrix}
        -H+\frac{\gamma}{c}I &0\\
        0 &\begin{bmatrix}H\\L\end{bmatrix}\!\!(H-\frac{1}{\gamma}\Phi^\top\Phi)^{-1}\!\!\begin{bmatrix}H\\L\end{bmatrix}^\top\!
    \end{bmatrix}+
\Pi_f(\tau)\prec 0,\label{M1:1}\\
&-H+\frac{1}{\gamma}\Phi^\top\Phi\prec  0.\label{M1:2}
\end{align}
\end{subequations}
According to \eqref{C}, given any $\tau$ satisfying constraint \eqref{sdp:robust_con3}, the inequality
\begin{equation}\label{C1}
    \begin{bmatrix}
I &A &B
\end{bmatrix}
\Pi_f(\tau)
\begin{bmatrix}
I &A &B
\end{bmatrix}^\top\succeq 0
\end{equation}
holds for any $(A, B)\in\mathcal{C}_f$.
Pre-multiplying \eqref{M1:1} with $\begin{bmatrix}I &A &B\end{bmatrix}$ and post-multiplying \eqref{M1:1} with $\begin{bmatrix}I &A &B\end{bmatrix}^\top$, the resulting inequality together with \eqref{C1} imply that the following inequality must hold for any $(A, B)\in\mathcal{C}_f$
\begin{equation}\label{X1}
    \begin{bmatrix}
        I \\
        A^\top\\
        B^\top
    \end{bmatrix}^\top
    \!\!\!
    \begin{bmatrix}\!
        -H+\frac{\gamma}{c}I \!\!\!\!&0\\
        0 \!\!\!\!&\begin{bmatrix}H\\L\end{bmatrix}\!(H-\frac{1}{\gamma}\Phi^\top\Phi)^{-1}\!\begin{bmatrix}H\\L\end{bmatrix}^\top\!
    \end{bmatrix}\!\!
    \begin{bmatrix}
        I\\
        A^\top\\
        B^\top
    \end{bmatrix}\!\!\prec 0.
\end{equation}
This is equivalent to
\begin{equation}\label{th1:1}
    (-H+\frac{\gamma}{c}I)+(AH+BL)(H-\frac{1}{\gamma}\Phi^\top\Phi)^{-1}(AH+BL)^\top\!\!\prec 0.
\end{equation}
Using the Schur complement, \eqref{th1:1} together with \eqref{M1:2} is equivalent to
\begin{equation}\label{th1:2}
    \begin{bmatrix}
        -H+\frac{1}{\gamma}\Phi^\top \Phi& (AH+BL)^\top\\
        (AH+BL) &-H+\frac{\gamma}{c} I
    \end{bmatrix}\prec  0.
\end{equation}
Using the Schur complement again, \eqref{th1:2} yields the equivalent inequality
\begin{subequations}\label{th1:3}
\begin{align}
    &(AH+BL)^\top (
    H\!-\!\frac{\gamma}{c} I)^{-1}(AH+BL)\!-\!H\!+\!\frac{1}{\gamma}\Phi^\top \Phi\prec  0,\label{th1:31}\\
    & -H+\frac{\gamma}{c} I\prec  0.\label{th1:32}
\end{align}
\end{subequations}
Let $P=\gamma H^{-1}$ and $F=LH^{-1}$.
Using the Woodbury matrix identity \cite{MR0038136}, we have $\gamma[P+P(cI-P)^{-1}P]^{-1}=H-\frac{\gamma}{c} I$.
Replacing $(H-\frac{\gamma}{c} I)^{-1}$ with  $\gamma^{-1}[P+P(cI-P)^{-1}P]$ in the inequality \eqref{th1:31}, multiplying both sides of the resulting inequality with $P$, and then dividing the resulting inequality by $\gamma$, we have
\begin{equation}\label{th1:4}
\begin{aligned}
    (A+BF)^\top [P+P(cI-P)^{-1}P](A+BF)-P\!\\
+Q+F^\top R F\prec  0.
\end{aligned}
\end{equation}
\addRed{Replacing $H$ with $\gamma P^{-1}$ in \eqref{th1:32} and dividing by $\frac{\gamma}{c}$, we have $-cP^{-1}+I\prec 0$.
Multiplying both sides of the resulting inequality with $P$, we obtain
\begin{equation}\label{th1:P}
    P^2-cP\prec 0.
\end{equation}
Using the Schur complement, \eqref{th1:P} is equivalent to
\begin{equation}\label{th1:P1}
\begin{bmatrix}
-cP &P\\
P &-I
\end{bmatrix}\prec 0.
\end{equation}
Using the Schur complement again and multiplying the resulting inequality with $c$, \eqref{th1:P1} is equivalent to
\begin{equation}\label{th1:P2}
P-cI\prec 0.
\end{equation}
}

Using the Schur complement, \eqref{th1:4} and \eqref{th1:P2} yield the equivalent inequality
\begin{equation}\label{th1:5}
    \begin{bmatrix}
    \!(A\!+\!BF)^\top \!P (A\!+\!BF)\!-\!P\!\!+\!\!Q\!\!+\!\!F^\top\! R F  &(A\!+\!BF)^\top P\\ P(A\!+\!BF) &P-cI
    \end{bmatrix}\!\!\prec  0.
\end{equation}
This implies for any $(A, B)\in\mathcal{C}_f$, we have
\begin{equation}\label{th1:51}
    (A+BF)^\top P (A+BF)-P+Q+F^\top R F\prec  0.
\end{equation}
Multiplying left and right sides of \eqref{th1:51} with $x^\top$ and $x$, the following inequality holds for any $x\in\mathbb{R}^n$ and any $(A, B)\in\mathcal{C}_f$
\begin{equation}\label{th1:6}
    x^\top \!(A+BF)^\top\! P(A+BF)x-x^\top \!P x\leq  -x^\top\! (Q+F^\top \!RF)x.
\end{equation}
The inequality \eqref{th1:6} implies that the following inequality is satisfied  for all  states and inputs $\bar{x}_k(t), \bar{u}_k(t)=F\bar{x}_k(t), k\in\mathbb{I}_{\geq 0}$ predicted by the system dynamics \eqref{mpc:robust_con1} with any $(A, B)\in\mathcal{C}_f$
\begin{equation}\label{Vconstraint}
\|\bar{x}_{k+1}(t)\|_{P}^2-\|\bar{x}_k(t)\|_{P}^2\leq -l(\bar{u}_k(t), \bar{x}_k(t)).
\end{equation}
Summing the inequality \eqref{Vconstraint} from $k=0$ to $k=T-1$ along an arbitrary trajectory, we obtain
\begin{equation}\label{Vsum}
\|\bar{x}_T(t)\|_{P}^2-\|\bar{x}_0(t)\|_{P}^2\leq -\sum_{k=0}^{T-1}l(\bar{u}_k(t), \bar{x}_k(t)).
\end{equation}
Since $\|\bar{x}_T(t)\|_{P}^2\geq 0$ and $\bar{x}_0(t)=x_t$, letting $T\to \infty$, we obtain
\begin{equation}
\label{SumV}
    \sum_{k=0}^{\infty}l(\bar{u}_k(t), \bar{x}_k(t))\leq \|x_t\|_{P}^2.
\end{equation}
The inequality \eqref{SumV} holds for any $(A, B)\in\mathcal{C}_f$, it also holds for the worst-case value, i.e., we obtain
\begin{equation}\label{worstV}
    \max_{(A, B)\in\mathcal{C}_f}\sum_{k=0}^{\infty}l(\bar{u}_k(t), \bar{x}_k(t))\leq \|x_t\|_{P}^2.
\end{equation}
This provides an upper bound on the optimal cost of \eqref{mpc:robust}.
Using the Schur complement, $\|x_t\|_{P}^2\leq \gamma$ is equivalent to the inequality \addRed{\eqref{sdp:robust_con1}}.

In conclusion, given that \eqref{sdp:robust_con1}-\eqref{sdp:robust_con3} hold, we have thus shown that $\gamma$ is an upper bound on the optimal cost of problem \eqref{mpc:robust} without input and state constraints \eqref{mpc:robust_con3}-\eqref{mpc:robust_con4}.
$\hfill\qed$
\end{pf}

\begin{remark}\upshape\label{remark:conser}
Theorem \ref{theorem1} derives an upper bound on the optimal cost of the data-driven min-max MPC problem \eqref{mpc:robust}.
Inequality \eqref{Vconstraint} together with the convergence of $\bar{x}_k(t)$ to the origin as $k$ tends to infinity allow us to show that $\|x_t\|_{P}^2$ serves as an upper bound on the infinite-horizon sum of stage costs of the nominal closed-loop system with any $(A, B)\in\mathcal{C}_f$.
The optimal solution of \eqref{sdp:robust} minimizes this upper bound and returns the corresponding state-feedback gain.
However, it is important to note that \addRed{the proposed method is conservative because }this upper bound may not always be tight.
\addRed{The SDP problem \eqref{sdp:robust} focuses on finding a linear state-feedback form of the input rather than a general input.
Besides, the upper bound is restricted to a quadratic form.}
Reducing conservatism by considering more general state-feedback law and cost upper bounding functions is an interesting issue for future research.
\end{remark}

\begin{remark}\upshape
The proof to derive the upper bound on the worst-case cost, i.e., \eqref{Vconstraint}-\eqref{worstV}, is inspired by the existing LMI-based min-max MPC approach in \cite{kothare1996robust}.
The difference is that \cite{kothare1996robust} propose a model-based min-max MPC scheme where the parametric uncertainty set is a predefined polytope.
On the other hand, in our case, $\mathcal{C}_f$ is an ellipsoidal set characterized by offline input-state trajectory generated by the noisy system, requiring different technical tools for the convex reformulation.
\end{remark}

\begin{remark}\upshape\label{remark:static}
In case of noise-free and persistently exciting data, i.e., \addRed{$G=\infty$} and $\begin{bmatrix}U_f^\top,X_f^\top\end{bmatrix}^\top$ has full row rank, the data-driven min-max MPC problem \eqref{mpc:robust} without the input and state constraints \eqref{mpc:robust_con3}-\eqref{mpc:robust_con4} reduces to a discrete-time linear quadratic regulator problem, as explored in \cite{van2020noisy,van2020data,berberich2023combining,persis2020formulas}.
In this case, the optimal state-feedback gain \addRed{$F_{x_t}^\star$} remains constant and does not depend on the state $x_t$.
However, in the presence of model uncertainty, even without the input and state constraints,
employing a receding horizon algorithm
and recalculating \addRed{$F_{x_t}^\star$} at each sampling time shows
significant performance improvement compared to using a static state-feedback control law.
This is illustrated with a numerical example in Section \ref{sec:5.1}.
\end{remark}

\subsection{Closed-loop Guarantees}\label{sec:3.3}

In the following theorem, we first establish the recursive feasibility of the problem \eqref{sdp:robust}.
Then, we \addRed{define a Lyapunov function $V(x)=\|x\|_{P_x^\star}^2$} and
prove robust stability for the resulting closed-loop system with any $(A, B)\in\mathcal{C}_f$.
Finally, we prove that the input and state constraints are satisfied for the closed-loop trajectory.

\begin{mythm}\label{theorem2}\upshape
\addRed{Suppose Assumption \ref{assumption1} holds and the optimization problem \eqref{sdp:robust} is feasible at time $t=0$ and $\gamma_{x_0}^\star\geq \frac{c^2}{\lambda_{\min}(Q)\lambda_{\min}(G)}$. Define $\mathcal{E}_{ROA}:=\{x\in\mathbb{R}^n:\|x\|_{P_{x_0}^\star}^2\leq\gamma_{x_0}^\star\}$ and $\mathcal{E}_{RPI}:=\{x\in\mathbb{R}^n: \|x\|_{\tilde{P}}^2\!\leq\! \frac{c^2}{\lambda_{\min}(Q)\lambda_{\min}(G)}\}$. Then, }
\begin{enumerate}[i)]
\item  the optimization problem \eqref{sdp:robust} is feasible \addRed{for any states $x_t\in\mathcal{E}_{ROA}\backslash\mathcal{E}_{RPI}$};
\item the set $\mathcal{E}_{RPI}$ is robustly stabilized for the closed-loop system \addRed{resulting from Algorithm~1} with any $(A, B)\in\mathcal{C}_f$;
\item the closed-loop trajectory \addRed{resulting from Algorithm~1} with any $(A, B)\in\mathcal{C}_f$ satisfies the constraints, i.e., $\|u_t\|_{S_u}\leq 1, \|x_t\|_{S_x}\leq 1$ for all $t\in \mathbb{I}_{\geq 0}$.
\end{enumerate}
\end{mythm}
\begin{pf}
The proof is composed of four parts.
Part I proves the lower bound and upper bound on the Lyapunov function $V(x_t)$.
Part II proves recursive feasibility of the problem \eqref{sdp:robust}.
Part III establishes the robust stability of the set $\mathcal{E}_{RPI}$ and Part IV shows that the input and state constraints are satisfied for the closed-loop trajectory.

\textbf{Part I: }First, we derive an upper bound and lower bound on $V(x_t)$.
As we have shown in Theorem \ref{theorem1}, $\|x_t\|_{P}^2$ with any feasible solution $P$ of the LMIs \eqref{sdp:robust_con1}-\eqref{sdp:robust_con3} is an upper bound on the optimal cost of \eqref{mpc:robust}.
Thus, $V(x_t)=\|x_t\|_{\addRed{{P_{x_t}^\star}}}^2$ is an upper bound on the optimal cost of \eqref{mpc:robust}.
Thus, we have
\begin{equation}\label{lowerbound}
    V(x_t)\geq l(u_t, x_t)\geq\|x_t\|_Q^2.
\end{equation}
In the proof of Theorem \ref{theorem1}, we have shown that any feasible solution of problem \eqref{sdp:robust} satisfies the inequality \eqref{th1:P2}.
Thus, $V(x_t)$ is upper bounded by $c\|x_t\|_2^2$.
Since $Q\succ 0$, we have $c\|x_t\|_2^2\leq \frac{c}{ \lambda_{\min}(Q)}\|x_t\|_Q^2$. Thus,
\begin{equation}\label{upperbound}
    V(x_t)\leq c\|x_t\|_2^2\leq \frac{c}{\lambda_{\min}(Q)}\|x_t\|_Q^2.
\end{equation}

\textbf{Part II: }
Assuming problem \eqref{sdp:robust} is feasible at time $t$, we have shown in the proof of Theorem \ref{theorem1} that the inequality \eqref{th1:5} holds for \addRed{$
F_{x_t}^\star, P_{x_t}^\star$} and any $(A, B)\in\mathcal{C}_f$.
Pre-multiplying \eqref{th1:5} with $\begin{bmatrix}x_t^\top \!\!&\omega_t^\top\end{bmatrix}$ and post-multiplying \eqref{th1:5} with $\begin{bmatrix}x_t^\top \!\!&\omega_t^\top\end{bmatrix}^\top$, we have
\begin{equation}\label{th2:5}
\begin{aligned}
    &[(A\!+\!B\addRed{F_{x_t}^\star})x_t\!+\!\omega_t]^\top \addRed{P_{x_t}^\star}[(A\!+\!B\addRed{F_{x_t}^\star})x_t\!+\!\omega_t]\!-\!x_t^\top \addRed{P_{x_t}^\star} x_t\\
    \leq &-x_t^\top (Q+\addRed{F_{x_t}^{\star\top}} R \addRed{F_{x_t}^\star})x_t+c\omega_t^\top \omega_t
\end{aligned}
\end{equation}
for any $x_t\in\mathbb{R}^n$, $\omega_t\in\mathbb{R}^n$ and $(A, B)\in\mathcal{C}_f$.
\addRed{Since $\omega_t$ satisfies the Assumption \ref{assumption1}, we have $\omega_t^\top \omega_t\leq\frac{1}{\lambda_{\min}(G)}\|\omega_t\|_G^2\leq \frac{1}{\lambda_{\min}(G)}$.}
Since $x_{t+1}=(A+BF_t^\star)x_t+\omega_t$, we have that
\begin{equation}\label{th2:51}
\begin{aligned}
\|x_{t+1}\|_{\addRed{{P_{x_t}^\star}}}^2\!-\!\|x_t\|_{\addRed{{P_{x_t}^\star}}}^2 &\!\leq\! -x_t^\top (Q+\addRed{F_{x_t}^{\star\top}} R \addRed{F_{x_t}^\star})x_t+c \omega_t^\top \omega_t\\
&\!\leq -\|x_t\|_Q^2+\addRed{\frac{c}{\lambda_{\min}(G)}}.
\end{aligned}
\end{equation}
By the upper bound on $V(x_t)=\|x_t\|_{\addRed{P_{x_t}^\star}}^2$ in \eqref{upperbound}, the inequality \eqref{th2:51} implies
\begin{equation}\label{th2:7}
\begin{aligned}
\|x_{t+1}\|_{\addRed{P_{x_t}^\star}}^2\!\!-\!\|x_t\|_{\addRed{P_{x_t}^\star}}^2 &\!\leq\!-\frac{\lambda_{\min}(Q)}{c}\|x_t\|_{\addRed{P_{x_t}^\star}}^2\!+\!\addRed{\frac{c}{\lambda_{\min}(G)}}.
\end{aligned}
\end{equation}
Subtracting $\addRed{\frac{c^2}{\lambda_{\min}(Q)\lambda_{\min}(G)}}$ from both sides of the inequality \eqref{th2:7} and adding $\|x_t\|_{\addRed{P_{x_t}^\star}}^2$ on both sides, we have
\begin{equation}\label{theorem2:e4}
\begin{aligned}
    &\|x_{t+1}\|_{\addRed{P_{x_t}^\star}}^2\!\!-\!\!\addRed{\frac{c^2}{\lambda_{\min}(Q)\lambda_{\min}(G)}}\!\\
    &\leq\!(1\!-\!\frac{\lambda_{\min}(Q)}{c}\!)(\|x_{t}\|_{\addRed{P_{x_t}^\star}}^2\!-\!\addRed{\frac{c^2}{\lambda_{\min}(Q)\lambda_{\min}(G)}}).
\end{aligned}
\end{equation}

As $c>\lambda_{\min}(Q)$, we derive $1>1-\frac{\lambda_{\min}(Q)}{c}>0$.
By the definition of $\mathcal{E}_{RPI}$ and $\mathcal{E}_{ROA}$, we have $V(x_t)=\|x_t\|_{\addRed{P_{x_t}^\star}}^2\in (\addRed{\frac{c^2}{\lambda_{\min}(Q)\lambda_{\min}(G)}}, \addRed{\gamma_{x_0}^\star}]$.
Thus, the inequality \eqref{theorem2:e4} implies
\begin{equation}\label{thm2:e5}
\|x_{t+1}\|_{\addRed{P_{x_t}^\star}}^2\!\!-\addRed{\frac{c^2}{\lambda_{\min}(Q)\lambda_{\min}(G)}}\!\!\leq\!\! \|x_{t}\|_{\addRed{P_{x_t}^\star}}^2\!\!-\addRed{\frac{c^2}{\lambda_{\min}(Q)\lambda_{\min}(G)}}.
\end{equation}
We further derive
\begin{equation}\label{thm2:ii1}
\|x_{t+1}\|_{\addRed{P_{x_t}^\star}}^2\leq \|x_{t}\|_{\addRed{P_{x_t}^\star}}^2\leq \gamma_t^\star.
\end{equation}
The only constraint in the problem \eqref{sdp:robust} that depends explicitly on the measured state $x_t$ is the inequality \eqref{sdp:robust_con1}.
The inequality \eqref{thm2:ii1} implies that the feasible solution of the optimization problem \eqref{sdp:robust} at time $t$ is also feasible at time $t+1$.
This argument can be continued to establish feasibility for any \addRed{state $x_t\in \mathcal{E}_{ROA}\backslash\mathcal{E}_{RPI}$}.

\textbf{Part III: }Now we separate the state $x_t$ into two cases to prove robust stability of the set $\mathcal{E}_{RPI}$.

\textbf{Case I:} \addRed{We first assume that $x_t \in \mathcal{E}_{ROA}\backslash\mathcal{E}_{RPI}$.}
Since \addRed{$P_{x_{t+1}}^\star=\gamma_{x_{t+1}}^\star (H_{x_{t+1}}^\star)^{-1}$} is the optimal solution of problem \eqref{sdp:robust} at time $t+1$ while \addRed{$P_{x_t}^\star$} is a feasible solution, we have
\begin{equation}\label{th2:optimal}
    V(x_{t+1})=\|x_{t+1}\|_{\addRed{P_{x_{t+1}}^\star}}^2\leq \|x_{t+1}\|_{\addRed{P_{x_t}^\star}}^2.
\end{equation}
When the state is outside the RPI set, i.e., $x_t\in \mathcal{E}_{ROA}\backslash\mathcal{E}_{RPI}$, we derive the following inequality using \eqref{theorem2:e4} and \eqref{th2:optimal}
\begin{equation}\label{theorem2:e5}
\begin{aligned}
    &V(x_{t+1})\!-\!\addRed{\frac{c^2}{\lambda_{\min}(Q)\lambda_{\min}(G)}}\!\\
    &\leq\!(1\!-\!\frac{\lambda_{\min}(Q)}{c}\!)[V(x_t)\!-\!\addRed{\frac{c^2}{\lambda_{\min}(Q)\lambda_{\min}(G)}}].
\end{aligned}
\end{equation}
Since $1>1-\frac{\lambda_{\min}(Q)}{c}>0$, the inequality \eqref{theorem2:e5} implies that all states in \addRed{$\mathcal{E}_{ROA}\backslash\mathcal{E}_{RPI}$} converge exponentially to the set $\mathcal{E}_{RPI}$.

\textbf{Case II:} \addRed{When the state is in the RPI set, i.e., $x_t\in\mathcal{E}_{RPI}$, we prove that the state stays inside the RPI set.}
\addRed{Since $\tilde{F}, \tilde{P}$ are derived from the optimal solution of problem \eqref{sdp:robust} at one time step during closed-loop operation, the inequality \eqref{th1:5} holds for $\tilde{F}, \tilde{P}$ and any $(A, B)\in\mathcal{C}_f$.
Following the same arguments as in \eqref{th2:5}-\eqref{thm2:e5}, we have}
\addRed{
\begin{equation}\label{thm2:p3c2}
\|x_{t+1}\|_{\tilde{P}}^2-\frac{c^2}{\lambda_{\min}(Q)\lambda_{\min}(G)}\leq
\|x_{t}\|_{\tilde{P}}^2-\frac{c^2}{\lambda_{\min}(Q)\lambda_{\min}(G)}.
\end{equation}}
By the definition of $\mathcal{E}_{RPI}$, we have $\|x_t\|_{\addRed{\tilde{P}}}^2\!\leq\! \addRed{\frac{c^2}{\lambda_{\min}(Q)\lambda_{\min}(G)}}$. The inequality \eqref{thm2:p3c2} implies that
\begin{equation}\label{theorem2:e6}
\|x_{t+1}\|_{\addRed{\tilde{P}}}^2\overset{}{\leq} \addRed{\frac{c^2}{\lambda_{\min}(Q)\lambda_{\min}(G)}}.
\end{equation}
which implies that the state $x_{t+1}$ stays inside the RPI set $\mathcal{E}_{RPI}$.
Thus, the set $\mathcal{E}_{RPI}$ is robustly stabilized for the closed-loop system \addRed{resulting from Algorithm~1} with any $(A, B)\in\mathcal{C}_f$.

\textbf{Part IV: }
Finally, we prove that the input and state constraints are satisfied for any closed-loop trajectory with $(A, B)\in\mathcal{C}_f$.

\addRed{\textbf{Case I:} We first assume that the state  $x_t \in \mathcal{E}_{ROA}\backslash\mathcal{E}_{RPI}$.} By constraint \eqref{sdp:robust_con1}, we have $x_t\in\mathcal{E}_t=\{x\in\mathbb{R}^n:\|x\|_{\addRed{P_{x_t}^\star}}^2\leq \addRed{\gamma_{x_t}^\star}\}$.
Given that the input is in a state-feedback form, we can write the input constraint \eqref{constraints:input} as
\begin{equation}\nonumber
\begin{aligned}
       \max_{t\in\mathbb{N}} \|u_t\|_{S_u}^2&=\max_{t\in\mathbb{N}} \|\addRed{F_{x_t}^\star} x_t\|_{S_u}^2\leq 1.
\end{aligned}
\end{equation}
For any state $x_t\in \mathcal{E}_t$, the state $x_{t+1}=(A+B\addRed{F_{x_t}^\star})x_t+\omega_t$ lies inside the set $\mathcal{E}_t$ at the next time step.
Thus, the input constraint \eqref{mpc:robust_con3} can be written as
\begin{equation}\label{theorem2:proof1}
\max\limits_{t\in\mathbb{N}} \|\addRed{F_{x_t}^\star} x_t\|_{S_u}^2\leq\max\limits_{x\in\mathcal{E}_t}\|\addRed{F_{x_t}^\star} x\|_{S_u}^2\leq 1.
\end{equation}
The inequality \eqref{theorem2:proof1} holds if
\begin{equation}\nonumber
    \begin{bmatrix}
        x\\
        1
    \end{bmatrix}^\top
    \begin{bmatrix}
        -\addRed{F_{x_t}^{\star\top}} S_u \addRed{F_{x_t}^\star}&0\\
        0 &1
    \end{bmatrix}
    \begin{bmatrix}
        x\\
        1
    \end{bmatrix}
    \geq 0,
\end{equation}
holds for all $x$ such that
\begin{equation}\nonumber
    \begin{bmatrix}
        x\\
        1
    \end{bmatrix}^\top
    \begin{bmatrix}
        -\addRed{P_{x_t}^\star} &0\\
        0 &\addRed{\gamma_{x_t}^\star}
    \end{bmatrix}
    \begin{bmatrix}
        x\\
        1
    \end{bmatrix}
    \geq 0.
\end{equation}
Using the S-procedure \cite{boyd1994linear}, if there exists $\addRed{\lambda_{x_t}}\geq 0$ such that
\begin{equation}\label{input:nominal_lmi1}
    \begin{bmatrix}
        -\addRed{F_{x_t}^{\star\top}} S_u \addRed{F_{x_t}^\star}&0\\
        0&1
    \end{bmatrix}-\addRed{\lambda_{x_t}}
    \begin{bmatrix}
        -\addRed{P_{x_t}^\star} &0\\
        0 &\addRed{\gamma_{x_t}^\star}
    \end{bmatrix}\succeq 0,
\end{equation}
then the input constraint \eqref{mpc:robust_con3} must be satisfied.
The inequality
\eqref{input:nominal_lmi1} holds iff the following inequalites hold
\begin{subequations}\label{input:nominal_lmi3}
    \begin{align}
        &\addRed{\lambda_{x_t} P_{x_t}^\star}-\addRed{F_{x_t}^{\star\top}} S_u \addRed{F_{x_t}^\star}\succeq 0,\label{input:nominal_lmi3a}\\
        &1-\addRed{\lambda_{x_t}\gamma_{x_t}^\star} \geq 0.\label{input:nominal_lmi3b}
    \end{align}
\end{subequations}
Without loss of generality, we choose the multiplier to be \addRed{$\lambda_{x_t}=\frac{1}{\gamma_{x_t}^\star}$}.
Multiplying both sides of \eqref{input:nominal_lmi3a} with \addRed{$H_{x_t}^\star$}, the inequality \eqref{input:nominal_lmi3a} is equivalent to
\begin{equation}\label{input:nominal_lmi4}
\addRed{H_{x_t}^\star}-\addRed{L_{x_t}^{\star\top}} S_u \addRed{L_{x_t}^\star}\succeq 0.
\end{equation}
Using the Schur complement, the inequality \eqref{input:nominal_lmi4} is equivalent to
\begin{equation}
    \begin{bmatrix}
        \addRed{H_{x_t}^\star} & \addRed{L_{x_t}^{\star\top}}\\
        \addRed{L_{x_t}^\star} &S_u^{-1}
    \end{bmatrix}\succeq 0.
\end{equation}
Similarly, the state constraint \eqref{mpc:robust_con4} can be written as \begin{equation}\nonumber
\begin{aligned}
       \max_{t\in\mathbb{N}} \|x_t\|_{S_x}^2\leq \max_{x\in\mathcal{E}_t}\|x\|_{S_x}^2\leq 1.
\end{aligned}
\end{equation}
Thus, the state constraint \eqref{mpc:robust_con4} holds if $x^\top S_x x\leq 1$ holds for all $x$ such that $x^\top \addRed{P_{x_t}^\star} x\leq\addRed{\gamma_{x_t}^\star}$.
The statement holds if $S_x\preceq \addRed{(\gamma_{x_t}^\star)^{-1} P_{x_t}^\star}$.
Multiplying \addRed{$H_{x_t}^\star$} from left and right, we have
\begin{equation}\label{state:robust1}
    \addRed{H_{x_t}^\star} S_x\addRed{H_{x_t}^\star}\preceq \addRed{H_{x_t}^\star}.
\end{equation}
Using the Schur complement, the inequality \eqref{state:robust1} is equivalent to \eqref{sdp:robust_con_state}.
\addRed{We have proved in Part II that the problem \eqref{sdp:robust} is recursively feasible for any state outside the RPI set. Thus, the closed-loop system with any $(A, B)\in\mathcal{C}_f$  satisfies the input and state constraints \eqref{constraints} in this case.}

\addRed{\textbf{Case II:} When the state $x_t$ is inside the RPI set, we have proved that the state stays inside the RPI set in Part III. Since the RPI set $\mathcal{E}_{RPI}$ is a subset of $\mathcal{E}_t$ in Case I and all states in $\mathcal{E}_t$ satisfy the state constraint, so the states inside the RPI set also satisfy the state constraint.}

\addRed{By Algorithm~1, if the state is inside the RPI set, the applied state-feedback gain is $\tilde{F}=F_{x_{t-i}}^\star$,
where $t-i$ is the last time step when $x_{t-i}\in\mathcal{E}_{ROA}\backslash\mathcal{E}_{RPI}$.
We know that $\|F_{x_{t-i}}^\star x\|_{S_u}^2\leq 1$ holds for any state $x$ satisfying $\|x\|_{P_{x_{t-i}}^\star}^2\leq\gamma_{t-i}^\star$. Since $\gamma_{x_{t-i}}^\star\geq \frac{c^2}{\lambda_{\min}(Q)\lambda_{\min}(G)}$, it follows that $\|\tilde{F} x\|_{S_u}^2\leq 1$ holds for any state $x$ satisfying $\|x\|_{\tilde{P}}^2\leq\frac{c^2}{\lambda_{\min}(Q)\lambda_{\min}(G)}$. Therefore, the input constraint holds for any state inside the RPI set.}

\addRed{Combining Case I and Case II, we have shown that the input and state constraints are satisfied for the closed-loop trajectory resulting from Algorithm~1.}

$\hfill\qed$
\end{pf}

\begin{remark}\upshape
Theorem \ref{theorem2} shows that if the optimization problem \eqref{sdp:robust} is feasible at initial time $t=0$ \addRed{and $\gamma_{x_0}^\star\geq\addRed{\frac{c^2}{\lambda_{\min}(Q)\lambda_{\min}(G)}}$}, then the closed-loop trajectory converges robustly and exponentially to the RPI set $\mathcal{E}_{RPI}$ for the closed-loop system with any $(A, B)\in\mathcal{C}_f$.
The idea is to construct a Lyapunov function $V$ satisfying the inequality \eqref{th2:51}.
The size of the RPI set depends on the pre-chosen constant $c$ and the \addRed{matrix $G$}.
\addRed{The Lyapunov function is defined using the optimal solution of problem \eqref{sdp:robust}, which is a state-dependent function. The RPI set is defined using $\tilde{P}$ resulting from Algorithm~1 and is a constant set.}
Furthermore, Theorem \ref{theorem2} shows that the input and state constraints are satisfied for the closed-loop system.
This is achieved via a Lyapunov function sublevel set $\mathcal{E}_t$ inside the constraints that contains the state.
Since $(A_s, B_s)\in\mathcal{C}_f$, the closed-loop trajectory of \eqref{system} is robustly stabilized and satisfy the input and state constaints.
\end{remark}


\begin{remark}\upshape
The length of the offline data influences the close-loop system performance.
A longer offline data sequence may lead to smaller volume of the set $\mathcal{C}_f$, hence leading to a better closed-loop performance.
However, this leads to a higher computational complexity.
As the length of offline input-state measurements $T_f$ increases, the computational complexity of problem \eqref{sdp:robust} increases due to the decision variable $\tau\in\mathbb{R}^{T_f}$.
It is possible to use different choices of multipliers which reduce the computational complexity at the cost of additional conservatism, see \cite[Section V.B]{berberich2023combining}. For example, when imposing the additional condition $\tau_i=\tau, \forall i \in\mathbb{I}_{[1, T_f]}$ for some $\tau\geq0$, the number of decision variables of problem \eqref{sdp:robust} is independent of the data length.
\end{remark}

\section{Adaptive Data-driven Min-Max MPC}\label{sec:4}
The data-driven min-max MPC scheme proposed in the previous section employs only an offline input-state data sequence to characterize consistent systems matrices for prediction.
During online operation,  the collection of additional online data can improve performance especially when offline data are inadequate.
To this end, in the present section, we propose an adaptive data-driven min-max MPC scheme using both offline and online data.

\subsection{Adaptive Data-driven Min-Max MPC Problem}\label{sec:4.1}
The online input-state measurements collected in the closed loop from initial time $0$ until time $t$ are denoted by
\begin{equation}\nonumber
    \begin{aligned}
        U_t&:=\begin{bmatrix}u_0 & u_1 &\ldots &u_{t-1}\end{bmatrix},\\
        X_t&:=\begin{bmatrix}x_0 &x_1 &\ldots &x_{t}\end{bmatrix}.
    \end{aligned}
\end{equation}
The set of system matrices $(A, B)$ consistent with the online data $x_t, u_t, x_{t+1}$ is defined by
\begin{equation}\nonumber
\Sigma_t^o\!:=\!\!\left\{\!(A, B)\!:\!
\eqref{system} \text{ holds for some }\omega_{t}
\text{ satisfying } \!\addRed{\|\omega_{t}\|_G}\!\leq\! 1
\right\}.
\end{equation} which can be characterized using the same manner as in equation \eqref{sigma_f}.
The set of system matrices consistent with the offline input-state sequence $(U_f, X_f)$ and the online input-state sequence $(U_t, X_t)$ is updated recursively by
\begin{equation}\label{update_C}
\mathcal{C}_{t}=
\begin{cases}
    \mathcal{C}_{t-1}\cap\Sigma_{t-1}^o,  &t\in\mathbb{I}_{[1,\infty)}\\
    \mathcal{C}_f, &t=0
\end{cases}
\end{equation}
where $\mathcal{C}_f$ is characterized by \eqref{C}.
The set $\mathcal{C}_t$ can also be written as
\[\mathcal{C}_{t}=\bigcap_{i=1}^{t-1}\Sigma_{i}^o\cap \bigcap_{i=0}^{T_f-1}\Sigma_{i}^f.\]
We can characterize $\mathcal{C}_t$ by the following inequality
\begin{equation}\label{C_t}
\mathcal{C}_t=\left\{\!\!(A, B)\!:\!\!\!
\begin{gathered}
\begin{bmatrix}
I &A &B
\end{bmatrix}
\!(\Pi_f(\tau)+\Pi_o(\delta))\!
\begin{bmatrix}
I &A &B
\end{bmatrix}^\top\!\succeq\! 0,  \\
\forall \tau\!=\!(\tau_0, \ldots, \tau_{T_f-1}), \tau_i\!\geq\! 0, i\in\mathbb{I}_{[0, T_f-1]}\\
\forall \delta\!=\!(\delta_0, \ldots, \delta_{t-1}), \delta_i\!\geq\! 0, i\in\mathbb{I}_{[0, t-1]}
\end{gathered}\!
\right\}\!,
\end{equation}
where $\Pi_f(\tau)$ is defined as in equation \eqref{Pi} using the offline input-state sequence and $\Pi_o(\delta)$ is defined by
\begin{equation}\label{Pi_o}
\Pi_o(\delta)=\sum_{i=0}^{t-1}\delta_i
\begin{bmatrix}
I &x_{i+1}^o\\
0 &-x_i^o\\
0 &-u_i^o
\end{bmatrix}
\!\!
\begin{bmatrix}
\addRed{G^{-1}} &0\\
0 &-I
\end{bmatrix}
\begin{bmatrix}
I &x_{i+1}^o\\
0 &-x_i^o\\
0 &-u_i^o
\end{bmatrix}^\top.
\end{equation}


Given the offline input-state sequence $(U_f, X_f)$ of length $T_f$, the online input-state sequence $(U_t, X_t)$, the current state $x_t$, the adaptive data-driven min-max MPC optimization problem is formulated as follows:
\begin{subequations}\label{mpc:adaptive}
\begin{align}
J_\infty^*(x_t):=&\min_{\bar{u}(t)}\max_{(A, B)\in\mathcal{C}_t}\sum_{k=0}^{\infty}l(\bar{u}_k(t), \bar{x}_k(t))\label{mpc:adaptive_obj}\\
\text{s.t. }\quad &\bar{x}_{k+1}(t)=A\bar{x}_k(t)+B\bar{u}_k(t),\label{mpc:adaptive_con1}\\
&\bar{x}_0(t)=x_t,\label{mpc:adaptive_con2}\\
&\addRed{\|\bar{u}_k(t)\|_{S_u}\leq 1, \forall k\in \mathbb{I}_{\geq 0},}\label{mpc:adaptive_con3}\\
&\addRed{\|\bar{x}_k(t)\|_{S_x}\leq 1, \forall (A, B)\in\mathcal{C}_t, k\in \mathbb{I}_{\geq 0},}\label{mpc:adaptive_con4}
\end{align}
\end{subequations}
Different from the data-driven min-max MPC problem \eqref{mpc:robust}, the objective function \eqref{mpc:adaptive_obj} is a minimization of the worst-case cost over all consistent system matrices in $\mathcal{C}_t$.
The set $\mathcal{C}_t$ is recursively updated using the input-state measurements collected online by \eqref{update_C}.
The state prediction, the initial state constraint and the input and state constraints remain the same as the data-driven min-max MPC problem.

\subsection{Reformulation based on LMIs}\label{sec:4.2}
To derive a tractable solution, we consider a state feedback control law $\bar{u}_k(t)=F_t\bar{x}_k(t)$ in problem \eqref{mpc:adaptive}, where $F_t\in\mathbb{R}^{m\times n}$ is the optimized state-feedback gain at time $t$.
In the following, we formulate an SDP to derive a state-feedback control law that minimizes an upper bound on the optimal cost of \eqref{mpc:adaptive} employing analogous techniques as in Section \ref{sec:3}.

At time $t\in\mathbb{I}_{[1, \infty)}$, given offline input-state measurements $(U_f, X_f)$ and online input-state measurements $(U_t, X_t)$, an initial state $x_t\in\mathbb{R}^n$ and a constant $c>\lambda_{\min}(Q)$, the SDP is formulated as follows:
\begin{subequations}\label{sdp:adaptive}
\begin{align}
    &\minimize\limits_{\gamma>0, H\in\mathbb{R}^{n\times n}, L\in\mathbb{R}^{m\times n}, \tau\in\mathbb{R}^{T_f}, \delta\in\mathbb{R}^{t}}\gamma\label{sdp:adaptive_obj}\\
    \text{s.t. }
    & \eqref{sdp:robust_con1}, \eqref{sdp:robust_con_input} \text{ and } \eqref{sdp:robust_con_state}\text{ hold},\label{sdp:adaptive_con1}\\
    &\begin{bmatrix}
        \begin{bmatrix}
            -H+\frac{\gamma}{c} I &0\\
            0 &0
        \end{bmatrix}+\Pi_f(\tau)+\Pi_o(\delta)\!\! &
        \begin{bmatrix}
            0\\
            H\\
            L
        \end{bmatrix}
        & 0\\
        \begin{bmatrix}
            0 &H &L^\top
        \end{bmatrix} \!\!&-H &\Phi^\top\\
        0 \!\!&\Phi & -\gamma I
    \end{bmatrix}\prec 0, \label{sdp:adaptive_con2}\\
    &\tau\!=\!(\tau_0, \ldots, \tau_{T_f-1}), \!\tau_i\geq 0, \forall i\in\mathbb{I}_{[0, T_f-1]},\label{sdp:adaptive_con3}\\
    &\delta\!=\!(\delta_0, \ldots, \delta_{t-1}), \delta_i\geq 0, \forall i\in\mathbb{I}_{[0, t-1]}.\label{sdp:adaptive_con4}
\end{align}
\end{subequations}
The optimal solution of the optimization problem \eqref{sdp:adaptive} \addRed{given state $x_t$} is denoted by $\addRed{\bar{\gamma}_{x_t}^\star, \bar{H}_{x_t}^\star, \bar{L}_{x_t}^\star, \bar{\tau}_{x_t}^\star, \bar{\delta}_{x_t}^\star}$, providing the optimal state-feedback gain \addRed{$\bar{F}_{x_t}^\star=\bar{L}_{x_t}^\star (\bar{H}_{x_t}^\star)^{-1}$}.

Similar to Theorem \ref{theorem1}, given the state $x_t$, we can show that the optimal cost of problem \eqref{mpc:adaptive} is guaranteed to be at most $\|x_t\|_{P}^2$ with $P=\gamma (H)^{-1}$ and $\|x_t\|_{P}^2$ is upper bounded by $\gamma$ if \eqref{sdp:robust_con1} and \eqref{sdp:adaptive_con2}-\eqref{sdp:adaptive_con3} hold for $\gamma, H, L, \tau, \delta$.
Therefore, problem \eqref{sdp:adaptive} minimizes an upper bound on the optimal cost of the adaptive data-driven min-max MPC problem \eqref{mpc:adaptive}.

We solve the SDP problem \eqref{sdp:adaptive} in a receding horizon manner, see Algorithm~2.
At time $t=0$, we solve the optimization problem \eqref{sdp:robust} and implement the first computed input $u_t=\addRed{\bar{F}_{x_t}^\star} x_t$.
At the next sampling time $t+1$, we measure the state $x_{t+1}$.
With the collection of online input-state data $x_t, u_t, x_{t+1}$, we adapt the set of consistent system matrices $\mathcal{C}_t$ by \eqref{C_t}.
A new optimization variable $\delta_t\geq 0$ is introduced to the optimization problem \eqref{sdp:adaptive} and the constraint \eqref{sdp:adaptive_con4} is updated.
Additionally, we update the the constraint \eqref{sdp:adaptive_con2} by incorporating the collected online input-state measurements $x_t, u_t, x_{t+1}$ and variable $\delta_t$ into $\Pi_o(\delta)$.
Then we solve the problem \eqref{sdp:adaptive}.
\addRed{Similar to Algorithm~1, we apply the first computed input $u_t=\bar{F}_{x_t}^\star x_t$ and repeat this procedure if $\bar{\gamma}_{x_t}^\star>\frac{c^2}{\lambda_{\min}(Q)\lambda_{\min}(G)}$.
If $\bar{\gamma}_{x_t}^\star\leq\frac{c^2}{\lambda_{\min}(Q)\lambda_{\min}(G)}$, we stop solving the problem \eqref{sdp:adaptive} and apply $\tilde{F}=\bar{F}_{x_{t-1}}^\star$.}

\begin{algorithm}[htb]
\begin{algorithmic}[1]
\caption{Adaptive data-driven min-max MPC scheme.}
    \State \algorithmicrequire{$U_f, X_f$, $Q, R, S_x, S_u$, $c, \addRed{G}$}\;
    \State At time $t=0$, measure state $x_0$\;
    \State Solve the problem \eqref{sdp:adaptive}\;
    \State Apply the input $u_t=\bar{F}_t^\star x_t$\;
    \State Set $t=t+1$ and measure state $x_t$\;
    \State Update the constraints \eqref{sdp:adaptive_con2} and \eqref{sdp:adaptive_con4}\;
    \State Solve the problem \eqref{sdp:adaptive}\;
    \addRed{\If{$\bar{\gamma}_t^\star>\frac{c^2}{\lambda_{\min}(Q)\lambda_{\min}(G)}$}\;
    \State Apply the input $u_t=\bar{F}_t^\star x_t$\;
    \State Set $t=t+1$,  measure state $x_t$, go back to 6\;
    \ElsIf{$\bar{\gamma}_t^\star\leq\frac{c^2}{\lambda_{\min}(Q)\lambda_{\min}(G)}$}
    \State Set $\tilde{F}=\bar{F}_{x_{t-1}}^\star, \tilde{P} = \bar{P}_{x_{t-1}}^\star$\;
    \State Apply the input $u_t=\tilde{F} x_t$\;
    \State Set $t=t+1$,  measure state $x_t$, go back to 13\;
    \EndIf}
    \label{algorithm:adaptive}
\end{algorithmic}
\end{algorithm}

In the following theorem, we show recursive feasibility of the problem \eqref{sdp:adaptive} and robust stability for the closed-loop system \eqref{system} resulting from the adaptive data-driven min-max MPC scheme.

\begin{mythm}\upshape\label{theorem3}
Suppose Assumption \ref{assumption1} holds and the optimization problem \eqref{sdp:adaptive} is feasible at time $t=0$ and $\addRed{\bar{\gamma}_{x_0}^\star}\geq \frac{c^2}{\lambda_{\min}(Q)\lambda_{\min}(G)}$. \addRed{Define $\mathcal{E}_{ROA}:=\{x\in\mathbb{R}^n:\|x\|_{\bar{P}_{x_0}^\star}^2\leq\bar{\gamma}_{x_0}^\star\}$ and $\mathcal{E}_{RPI}:=\{x\in\mathbb{R}^n: \|x\|_{\tilde{P}}^2\!\leq\! \frac{c^2}{\lambda_{\min}(Q)\lambda_{\min}(G)}\}$.}
Then,
\begin{enumerate}[i)]
\item the optimization problem \eqref{sdp:adaptive} is feasible \addRed{for any states $x_t\in\mathcal{E}_{ROA}\backslash\mathcal{E}_{RPI}$};
\item the set $\mathcal{E}_{RPI}$  is robustly stabilized for the closed-loop system \eqref{system} resulting from Algorithm 2;
\item the closed-loop trajectory of \eqref{system} resulting from Algorithm 2 satisfies the constraints, i.e., $\|u_t\|_{S_u}\leq 1, \|x_t\|_{S_x}\leq 1$ for all $t\in \mathbb{I}_{\geq 0}$.
\end{enumerate}
\end{mythm}
\begin{pf}
The proof is similar to that of Theorem \ref{theorem2}; hence, we only provide a sketch.
The difference is that a new optimization variable is introduced to the problem \eqref{sdp:adaptive} at each time step.
To prove recursive feasibility, suppose the optimal solution of the problem \eqref{sdp:adaptive} at time $t$ is \addRed{$\bar{\gamma}_{x_t}^\star, \bar{H}_{x_t}^\star, \bar{L}_{x_t}^\star, \bar{\tau}_{x_t}^\star, \bar{\delta}_{x_t}^\star$}.
When the state is outside the RPI set, i.e., $x_t\in\mathcal{E}_{ROA}\backslash\mathcal{E}_{RPI}$, we define a candidate solution of problem \eqref{sdp:adaptive} at time $t+1$ as follows:
\addRed{
\begin{equation}\nonumber
\begin{aligned}
&\bar{\gamma}'_{x_{t+1}}\!=\!\bar{\gamma}_{x_t}^\star, \bar{H}'_{x_{t+1}}\!=\!\bar{H}_{x_t}^\star, \bar{L}'_{x_{t+1}}\!=\!\bar{L}_{x_t}^\star,\\
&\bar{\tau}_{x_{t+1}}'\!=\!
\bar{\tau}_{x_t}^\star, \bar{\delta}'_{x_{t+1}}\!=\!\begin{bmatrix}\bar{\delta}_{x_t}^\star,0\end{bmatrix}.
\end{aligned}
\end{equation}}
Constraint \eqref{sdp:adaptive_con2} is trivially satisfies with the defined candidate solution.

The satisfaction of \eqref{sdp:robust_con1} and \eqref{sdp:adaptive_con2}-\eqref{sdp:adaptive_con4} implies that \eqref{th1:5} is satisfied for any $(A, B)\in\mathcal{C}_t$.
The set $\mathcal{C}_t$ is recursively updated over time.
Based on its definition, the uncertainty set $\mathcal{C}_t$ satisfies $(A_s, B_s)\in\mathcal{C}_{t+1}\subseteq\mathcal{C}_t$ for all $t\in\mathbb{N}$.
As $t$ approaches infinity, we can establish that \eqref{th1:5} holds for any $(A, B)\in\mathcal{C}_t$, where the true system matrices always lie within this set.
Using \eqref{th1:5}, we can show robust stability and constraint satisfaction for the closed-loop system \eqref{system} following the same steps as Theorem \ref{theorem2}.
$\hfill\qed$
\end{pf}

\begin{remark}\upshape
As the online input-state measurements are collected and $\mathcal{C}_t$ is recursively updated, the uncertainty due to the set $\mathcal{C}_t$ never \addRed{increases} and may possibly \addRed{decrease}.
Thus, the closed-loop performance resulting from the adaptive data-driven min-max MPC scheme is no worse than the approach in Section \ref{sec:3}.
\addRed{It is an interesting future direction to investigate conditions under which we can guarantee a decrease in uncertainty with online measurements.
One promising approach involves leveraging the concept of persistently exciting condition, compare \cite{lu2021robust}.}
\end{remark}

\addRed{\begin{remark}\upshape
While the proposed data-driven min-max MPC addresses LTI systems, our results open up promising directions for addressing nonlinear systems as a next step.
In particular, it may be possible to handle nonlinear systems by exploiting online measurements to characterize the consistent system matrices for the linearized system dynamics around the current
operating point.
As the system dynamics evolve, previously collected offline data and outdated online data
may become less relevant or even misleading for the current system dynamics.
In this case, it may be necessary to forget some offline data and previously collected online data to accurately characterize the current system dynamics, see \cite{berberich2022linear} for analogous results on data-driven MPC based on the Fundamental Lemma.
Addressing this issue is beyond the scope of this paper but is an interesting direction for future research.
\end{remark}}

\begin{remark}\upshape
The adaptive data-driven min-max MPC scheme directly incorporates new data to the SDP problem \eqref{sdp:adaptive}, which increases the computational complexity due to the introduction of a new optimization variable $\delta_t$ in problem \eqref{sdp:adaptive} at each time step.
One possible method to reduce the computational complexity is to stop adding new data when the closed-loop performance is satisfactory.
Alternatively, one can
recursively compute an outer approximation of the set $\mathcal{C}_t$ using the previous uncertainty set along with the newly collected data at time $t$.
Possible methods for computing an outer approximation have been investigated in \cite{martin2023inference,bisoffi2021trade}.
\end{remark}

\section{Simulation}\label{sec:5}

In this section, we demonstrate the effectiveness of the proposed data-driven min-max MPC schemes through two numerical examples.
First, we implement the proposed MPC schemes in Sections \ref{sec:3} and \ref{sec:4} on \addRed{an active suspension system}.
Both schemes have good closed-loop performance and the adaptive scheme shows performance improvement compared to the approach in Sections \ref{sec:3}.
Second, we compare the proposed approach with the robust data-driven MPC scheme based on the Fundamental Lemma from \cite{berberich2020robustcon}.
Simulation results illustrate that our scheme is less conservative and achieves better closed-loop performance.

\subsection{Implementation of the Proposed MPC Scheme}\label{sec:5.1}
\addRed{We consider an active suspension system \cite{hrovat1997survey}, the state-space equation is given by
\begin{equation}\label{system:LTI}
x_{t+1}\!=\!\begin{bmatrix}
    0.809 &0.009 &0 &0\\
    -36.93 &0.8 &0 &0\\
    0.191 &-0.009 &1 &0.01\\
    0 &0 &0 &1
\end{bmatrix}\!x_t\!+\!\begin{bmatrix}
    0.0005 \\ 0.0935 \\ -0.005 \\ -0.01
\end{bmatrix}\!u_t\!+\!\omega_t,
\end{equation}}
where the noise satisfies Assumption~\ref{assumption1} \addRed{ with
$G=1\times 10^{8}I$.}
The system matrices are unknown, but an offline input-state trajectory $(U_f, X_f)$ of length $T_f=\addRed{200}$ is available, where the input $u\in U_f$ is chosen uniformly from the unit interval \addRed{$[-5, 5]$}.
We choose the weighting matrices of the stage cost function by \addRed{$Q=100I, R=1$}.
Moreover, the input and state constraints are given by $\|u_t\|_{S_u}\leq 1$ and $\|x_t\|_{S_x}\leq 1$, where \addRed{\[S_u=0.25, S_x=\begin{bmatrix}2500 &0 &0 &0\\0 &1 &0 &0\\ 0 &0 &400 &0\\ 0 &0 &0 &1\end{bmatrix}.\]}

\begin{figure}
    \centering
    \includegraphics[width=0.5\textwidth]{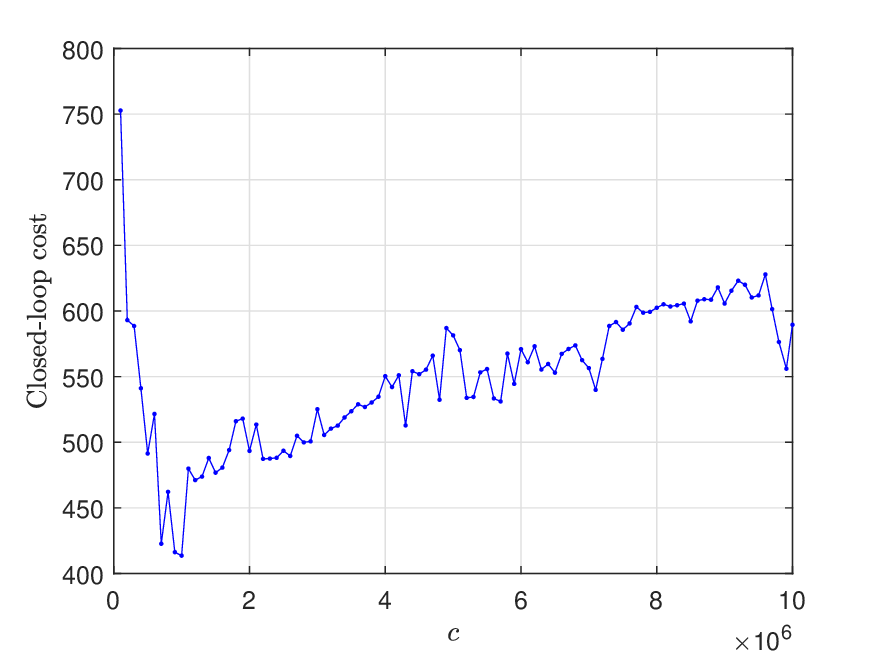}
    \caption{\addRed{Closed-loop cost over $150$ iterations under the proposed data-driven min-max MPC scheme with different values of $c$.}}
    \label{pic:compare_c}
\end{figure}

\addRed{We implement the data-driven min-max MPC scheme in Section \ref{sec:3} with different values of $c$.}
The initial state is given as \addRed{$x_0=[-0.01,-0.5,0.03,0.1]^\top$}.
Figure \ref{pic:compare_c} illustrates \addRed{the closed-loop cost over $150$ iterations under} the proposed data-driven min-max MPC scheme in Section~\ref{sec:3} \addRed{with different values of $c$ in the range $[5\times 10^4, 1\times 10^7]$}.
All closed-loop state trajectories converge to a neighborhood of the origin \addRed{with $c$ in this range}.
The input and state constraints are satisfied in the closed-loop operation.
\addRed{When $c$ is too small, i.e., $c=100$, the SDP problem \eqref{sdp:robust} is infeasible at the initial time.
When $c=5\times 10^4$, the feasible region is small which leads to poor closed-loop performance.
On the other hand, when $c$ is too large, i.e., $c=1\times 10^7$, the system starts inside the RPI set and results in static state-feedback gain and worse performance.}
\addRed{
To achieve good closed-loop performance, $c$ can be initially set to a large value and gradually reduced until satisfactory performance is reached.
Note that a good choice of $c$ depends on the stage cost matrix $Q$ and the noise level $G$. Thus, $c$ should be chosen in conjunction with these parameters to optimize performance.}


\addRed{We now compare the robust data-driven min-max MPC scheme from Section~\ref{sec:3} to the adaptive approach from Section~\ref{sec:4}.
We consider an offline input-state trajectory of length $T_f=150$ and set $c=5\times 10^5$.}
Figure 2 illustrates the closed-loop input trajectories resulting from the application of the proposed
data-driven min-max MPC schemes and the static statefeedback control law.
The static state-feedback gain is
computed at time $t = 0$ of the data-driven min-max MPC scheme in Section 3, as explained in Remark \ref{remark:static}.
The closed-loop system resulting from these three schemes converges to the RPI set.
\addRed{The resulting closed-loop state of the data-driven min-max MPC scheme in Section \ref{sec:3} converges to the RPI set after 7 iterations, whereas the resulting closed-loop state of the data-driven min-max MPC scheme in Section \ref{sec:4} converges to the RPI set after 2 iterations.}
Table \ref{table:LTI} presents the the sum of closed-loop stage costs over all \addRed{150} iterations and the average computational time per iteration for the proposed data-driven min-max MPC schemes and the static state-feedback control law.
The closed-loop cost of the adaptive data-driven min-max MPC scheme is \addRed{$29.81\%$} smaller than
the MPC scheme in Section \ref{sec:3} and the average computational time per
iteration is larger than the scheme in Section \ref{sec:3}.

\begin{figure}
    \centering
    \includegraphics[width=0.5\textwidth]{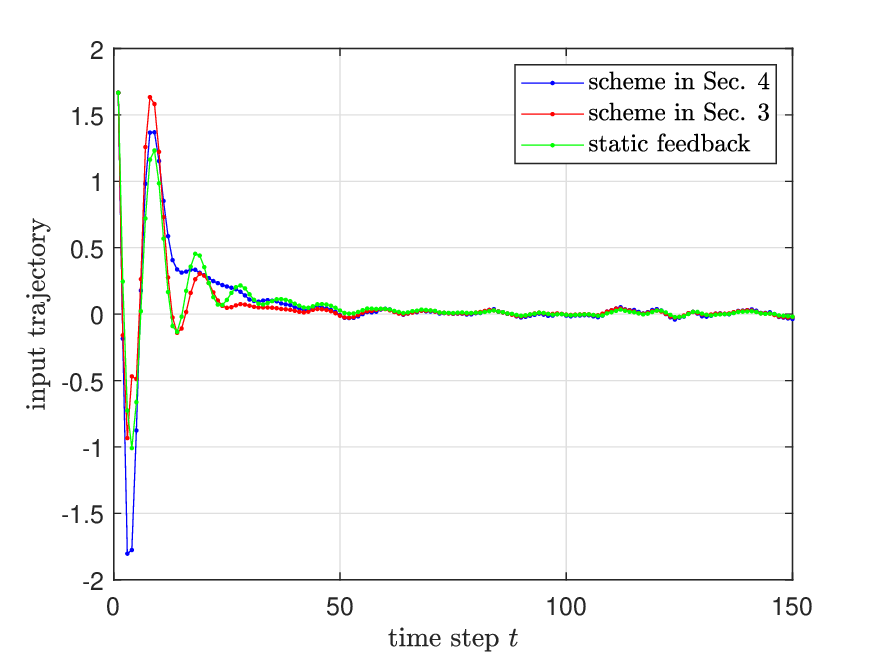}
    \caption{Closed-loop input trajectory under the proposed data-driven min-max MPC schemes.}
    \label{pic:input}
\end{figure}

\addRed{
\begin{table}[ht]
\centering
\caption{Closed-loop cost and average computation time.}
\label{table:LTI}
\begin{tabular}{ c|c|c}
\hline
\hline
&cost &time (s)\\
\hline
static state-feedback &\addRed{766.83} &/\\
\hline
scheme in Sec.3&\addRed{684.90} &\addRed{0.2613}\\
\hline
scheme in Sec.4&\addRed{480.75}
 &\addRed{0.3990}\\
\hline
\hline
\end{tabular}
\end{table}
}


\subsection{Comparison to the Data-driven MPC Schemes in literature}\label{sec:5.2}
In this section, we contrast our data-driven min-max MPC scheme with the data-driven MPC scheme from \cite{berberich2020robustcon}, which relies on the Fundamental Lemma and includes a constraint tightening guaranteeing robust constraint satisfaction.
While our approach incorporates ellipsoidal constraints, \cite{berberich2020robustcon} employs hypercube constraints.
To facilitate a comparative analysis, we implement both schemes on a scalar system
\begin{equation}\label{system_numerical2}
    x_{t+1}=1.1x_t+0.5u_t+\omega_t,
\end{equation}
where the noise satisfies $\omega_t\in\{\omega\in\mathbb{R}:\addRed{\|\omega\|_G\leq 1\}}$ with \addRed{$G=10^{8}$}.
The input and state constraints are $|u_t|\leq 2$ and $|x_t|\leq 2$.
An input-state trajectory $(U_f, X_f)$ of length $T_f=20$ is available.
The weighting matrices of the stage cost function are $Q=1, R=0.1$.
The initial state is given as $x_0=-1$.

\begin{figure}
    \centering
    \subfigure[]{\label{pic:state_lit}
    \includegraphics[width=0.51\textwidth]{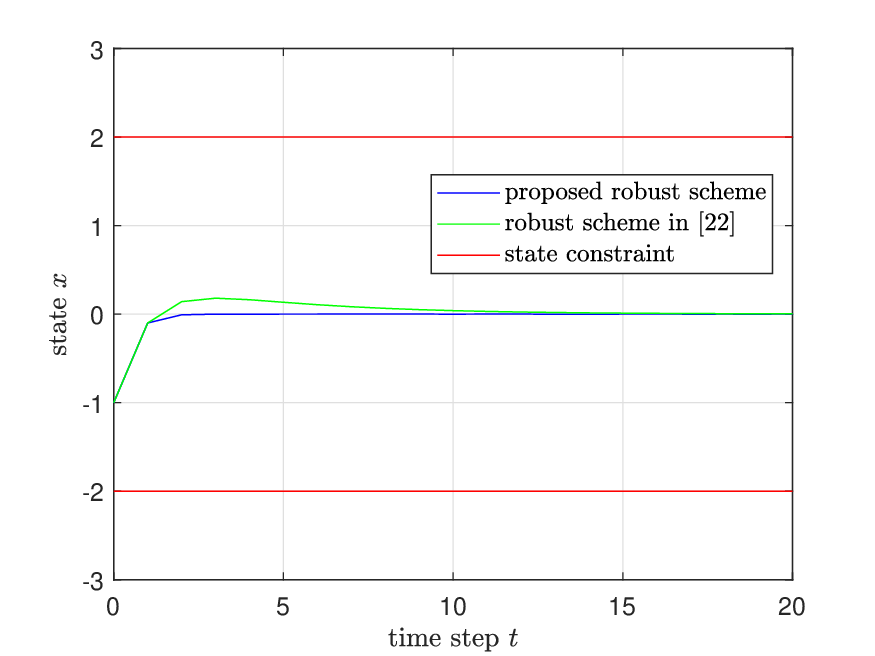}}
    \subfigure[]{\label{pic:input_lit}
    \includegraphics[width=0.51\textwidth]{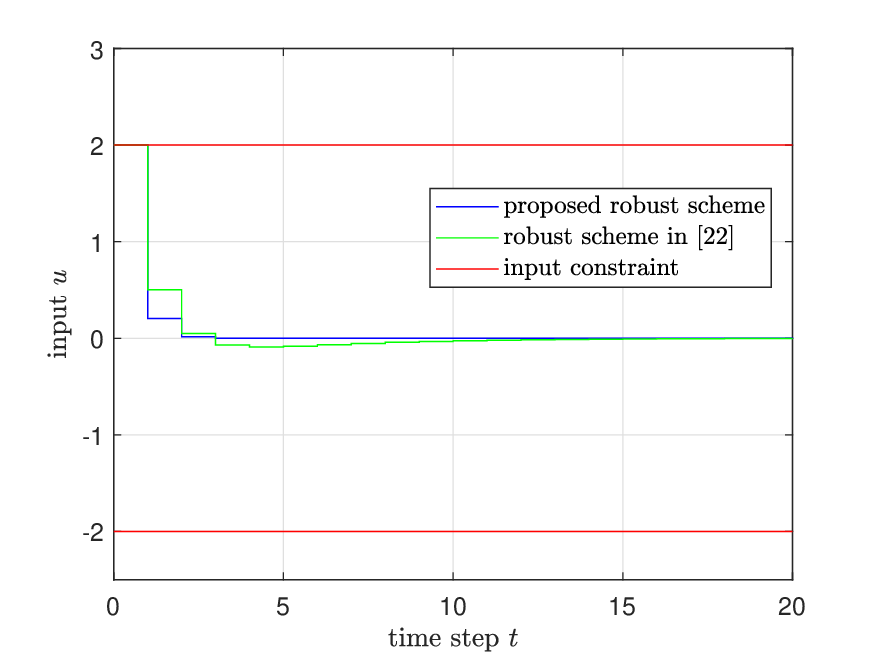}}
    \subfigure[]{\label{pic:cost}
    \includegraphics[width=0.51\textwidth]{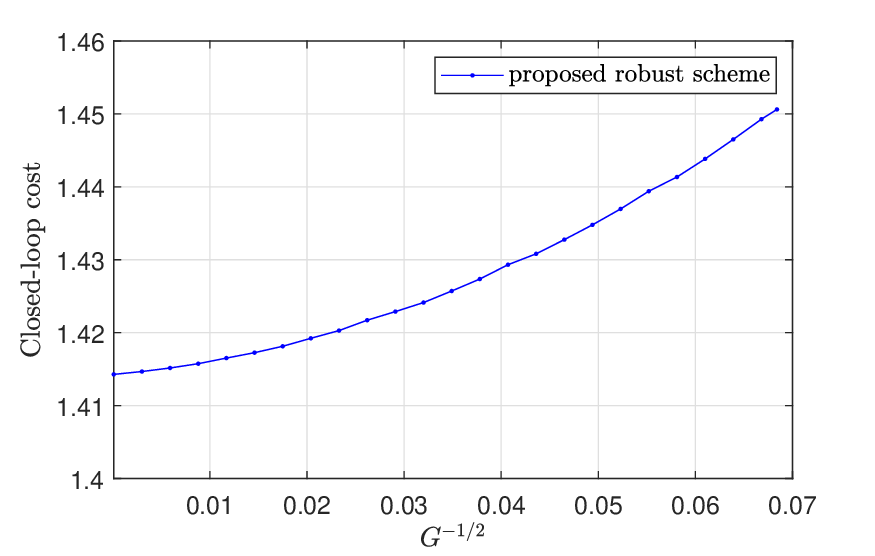}}
    \caption{\addRed{(a)-(b) Closed loop state and input trajectories under the proposed data-driven min-max MPC scheme and the robust constraint tightening data-driven MPC scheme in [22]; (c) Closed-loop cost of the proposed data-driven min-max MPC scheme with different noise levels.}}
    \label{pic:compare_lit}
\end{figure}

We apply the proposed data-driven min-max MPC scheme and the data-driven MPC scheme in \cite{berberich2020robustcon}.
While our approach accounts for process noise in the system described by \eqref{system_numerical2}, \cite{berberich2020robustcon} focuses on measurement noise.
To translate the bound on process noise into a bound on measurement noise as required  for \cite{berberich2020robustcon}, we use the fact that process noise bounded by \addRed{$\epsilon=G^{-\frac{1}{2}}$} results in measurement noise bounded by $\sum_{i=0}^{k-1}A_s^i\epsilon$ at time $k$.
Figure \ref{pic:compare_lit}\addRed{(a)-(b)} illustrates the closed-loop input and state trajectories resulting from the application of both schemes.
The input and state trajectories from both schemes converges to a neighborhood of the origin and satisfy the input and state constraints.
The sum of closed-loop stage
costs over all 20 iterations for the proposed data-driven min-max MPC scheme is $9.58\%$ lower than that for data-driven MPC scheme in \cite{berberich2020robustcon}.
\addRed{The average computation time per iteration for the proposed data-driven min-max MPC scheme is 0.058 seconds, compared to 0.1171 seconds for the data-driven MPC scheme in \cite{berberich2020robustcon}.}

We now increase the bound on the noise and implement both schemes as explained above.
\addRed{Figure~3(c) illustrates the closed-loop cost of the proposed data-driven min-max MPC scheme with different noise levels.}
When \addRed{$G^{-\frac{1}{2}}$} approaches \addRed{$0.00012$}, the approach from \cite{berberich2020robustcon} becomes infeasible.
In contrast, the proposed MPC scheme remains feasible and robustly stabilizes the system for \addRed{$G^{-\frac{1}{2}}$} up to \addRed{$0.0684$}.
\addRed{The closed-loop cost of the scheme in \cite{berberich2020robustcon} is $2.0670$ when $G^{-\frac{1}{2}}=0.00012$, which is much larger than that of the proposed data-driven min-max MPC scheme.}
This result shows that our proposed data-driven min-max MPC scheme exhibits less conservatism compared to the approach in \cite{berberich2020robustcon}, allowing for stability and constraint satisfaction guarantees with higher noise levels.
Further, as shown in Section \ref{sec:4}, it allows to employ online data in order to improve closed-loop performance, which is not easily possible for the approach from \cite{berberich2020robustcon}.

\section{Conclusion}\label{sec:6}
In this paper, we present a data-driven min-max MPC scheme that uses noisy input-state data to design state-feedback controllers for unknown LTI systems.
We reformulate the data-driven min-max MPC problem with ellipsoidal input and state constraints as an SDP.
A receding-horizon algorithm is proposed to repeatedly solve the SDP at each time step and obtain a state-feedback gain.
We establish that the proposed scheme guarantees closed-loop  recursive feasibility, constraint satisfaction and robust stability for any systems consistent with the noisy input-state data.
Furthermore, we propose an adaptive data-driven min-max MPC scheme that employing online collected input-state data to improve closed-loop performance when the offline data are insufficient.
We establish that the resulting closed-loop trajectory satisfies the input and state constraint and is robustly stabilized.
Two numerical examples show that the adaptive scheme indeed improve the closed-loop performance compared with the robust scheme, and our proposed data-driven min-max MPC scheme exhibits less conservatism than the robust constraint tightening MPC scheme in the literature
In the future, we plan to investigate the data-driven min-max MPC scheme using noisy input-output data.
Further, extending our results to nonlinear systems is another interesting direction.

\bibliographystyle{unsrt}
\bibliography{main}

\end{document}